\documentclass[12pt]{article}
\usepackage{epsf}
\usepackage{amsmath}
\usepackage{graphics}

\renewcommand{\thefootnote}{\#\arabic{footnote}}

\begin{document}

\setcounter{footnote}{0}
\begin{titlepage}

\begin{center}

\hfill astro-ph/0601099\\
\hfill January 2006\\

\vskip .5in

{\Large \bf
Reexamining the Constraint on the Helium  Abundance from CMB 
}

\vskip .45in

{\large
Kazuhide Ichikawa and Tomo Takahashi
}

\vskip .45in

{\em
Institute for Cosmic Ray Research,
University of Tokyo\\
Kashiwa 277-8582, Japan
}

{\it (\today)}

\end{center}

\vskip .4in

\begin{abstract}

We revisit the constraint on the primordial helium mass fraction $Y_p$
from observations of cosmic microwave background (CMB) alone. By
minimizing $\chi^2$ of recent CMB experiments over 6 other cosmological
parameters, we obtained rather weak constraints as $0.17\le Y_p \le
0.52$ at 1$\sigma$ C.L. for a particular data set.  We also study the
future constraint on cosmological parameters when we take account of the
prediction of the standard big bang nucleosynthesis (BBN) theory as a
prior on the helium mass fraction where $Y_p$ can be fixed for a given
energy density of baryon. We discuss the implications of the prediction
of the standard BBN on the analysis of CMB.
\vspace{1cm}
\end{abstract}

\end{titlepage}

\renewcommand{\thepage}{\arabic{page}}
\setcounter{page}{1}
\renewcommand{\thefootnote}{\#\arabic{footnote}}
\renewcommand{\theequation}{\thesection.\arabic{equation}}

\section{Introduction}
\setcounter{equation}{0}

Recent precise cosmological observations such as WMAP
\cite{Bennett:2003bz} push us toward the era of so-called precision
cosmology.  In particular, the combination of the data from cosmic
microwave background (CMB), large scale structure, type Ia supernovae
and so on can severely constrain the cosmological parameters such as the
energy density of baryon, cold dark matter and dark energy, the equation
of state for dark energy, the Hubble parameter, the amplitude and the
scale dependence of primordial fluctuation.

Among the various cosmological parameters, the primordial helium mass
fraction $Y_p$ is the one which has been mainly discussed in the context
of big bang nucleosynthesis (BBN) but not that of CMB so far. One of the
reason is that the primordial helium abundance has not been considered
to be well constrained by observations of CMB since its effects on the
CMB power spectrum is expected to be too small to be measured.  However,
since now we have very precise measurements of CMB,  we may have a chance
to constrain the primordial helium mass fraction from CMB
observations. Since the primordial helium mass fraction can affect the
number density of free electron in the course of the recombination
history, the effects of $Y_p$ can be imprinted on the power spectrum of
CMB. Recently, some works along this line have been done by two
different groups \cite{Trotta:2003xg,Huey:2003ef}, which have discussed
the constraints on $Y_p$ from current observations of CMB.  In fact they
claim different bounds on the primordial helium mass fraction,
especially in terms of its uncertainty: the author of
Ref.~\cite{Trotta:2003xg} obtained $ 0.160 \le Y_p \le 0.501$, on the
other hand the authors of Ref.~\cite{Huey:2003ef} got $ Y_p
=0.250^{+0.010}_{-0.014}$ at 1$\sigma$ confidence level.  It should be
noticed that the latter bound is much more severe than that of the
former. If the helium mass fraction is severely constrained by CMB data,
it means that the CMB power spectrum is sensitive to the values of
$Y_p$.  In such a case, the prior on $Y_p$ should be important to
constrain other cosmological parameters too and the usual fixing of
$Y_p=0.24$ in CMB power spectrum calculations might not be a good
assumption. Especially, analyses like
Refs.~\cite{Cyburt:2003fe,Cuoco:2003cu,Coc:2003ce,Cyburt:2004cq,Serpico:2004gx}
predict light element abundances including $^4$He from the baryon
density which is obtained from the CMB data sets with the analysis fixing the value of
$Y_p$. Such procedure is only valid when $Y_p$ is not severely
constrained by CMB. Thus it is very important to check the CMB bound on
$Y_p$.

One of the main purpose of the present paper is that we revisit the
constraint on $Y_p$ from observations of CMB alone with a different
analysis method from Markov chain Monte Carlo (MCMC) technique which is
widely used for the determination of cosmological parameters and adopted
in Refs.~\cite{Trotta:2003xg,Huey:2003ef}.  In this paper, we calculate
$\chi^2$ minimum as a function of $Y_p$ and derive constraints on
$Y_p$. We adopt the Brent method of the successive parabolic
interpolation to minimize $\chi^2$ varying 6 other cosmological
parameters of the $\Lambda$CDM model with the power-law adiabatic
primordial fluctuation. We obtain the constraint on $Y_p$ by this method
and compare it with previously obtained results.

We also study the constraint on $Y_p$ from future CMB experiment. A
particular emphasis is placed on investigating the role of the standard
BBN theory. Since the primordial helium is synthesized in BBN, once the
baryon-to-photon ratio is given, the value of $Y_p$ is fixed
theoretically. Thus, using this relation between the baryon density and
helium abundance, we do not have to regard $Y_p$ as an independent free
parameter when we analyze CMB data. We study how the standard BBN
assumption on $Y_p$ affects the determination of other cosmological
parameters in the future Planck experiment using the Fisher matrix
analysis.

The structure of this paper is as follows. In the next section, we
briefly discuss the effects of the helium mass fraction on the CMB power
spectrum, in particular its effects on the change of the structure of
the acoustic peaks. Then we study the constraint on the primordial
helium mass fraction from current observations of CMB using the data
from WMAP, CBI, ACBAR and BOOMERANG. In section 4, we discuss the
expected constraint on $Y_p$ from future CMB observation of Planck and
also study how the standard BBN assumption on $Y_p$ can affect the
constraints on cosmological parameters. The final section is devoted to
the summary of this paper.

\section{Effects of the change of $Y_p$ on CMB}

In this section, we briefly discuss the effects of the change of the
helium abundance on the CMB power spectrum. More detailed description of
this issue can be found in Ref.~\cite{Trotta:2003xg}.

\begin{figure}[t]
\begin{center}
\scalebox{1}{\includegraphics{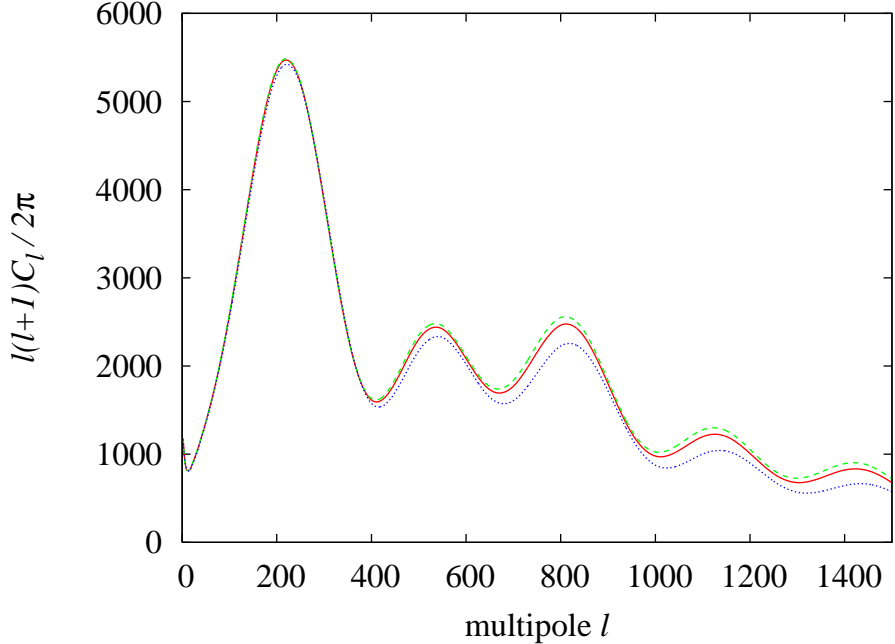}}
\caption{The CMB power spectra for the cases with $Y_p=0.1$ (green dashed
line), $0.24$ (red solid line) and $0.5$ (blue dotted
line). Other cosmological parameters are taken to be the WMAP mean values
for the power-law $\Lambda$CDM model.}
\label{fig:cl}
\end{center}
\end{figure}

The main effect of $Y_p$ on the CMB power spectrum comes from the
diffusion damping at small scales. When $Y_p$ is large, since it is
easier for electrons to recombine with $^4$He than with H, the number of
free electron becomes small. Thus the Compton mean free path becomes
larger for larger $Y_p$, which means that the diffusion length of photon
becomes also larger. Since the photon-baryon tight coupling breaks down
at the photon diffusion scales, the fluctuation of photon is
exponentially damped due to the diffusive mixing and rescattering.
Hence the CMB power spectrum is more damped for larger values of $Y_p$.
To see this tendency, we plot the CMB power spectra for several values
of $Y_p$ in Fig.~\ref{fig:cl}. We clearly see that the damping at the
small scales is more significant for the cases with larger values of
$Y_p$. The effect of diffusion damping causes the change in the power
spectrum at a percent level for 10 \% change of $Y_p$
\cite{Trotta:2003xg}.

To see this more quantitatively, we consider the ratio of the second
peak height to the first which is defined as \cite{Hu:2000ti}
\begin{eqnarray}
H_2 \equiv \left( \frac{\Delta T(l=l_2)}{\Delta T(l=l_1)} \right)^2,
\end{eqnarray}
and the third peak height to the first
\begin{eqnarray}
H_3 \equiv \left( \frac{\Delta T(l=l_3)}{\Delta T(l=l_1)} \right)^2,
\end{eqnarray}
where $(\Delta T(l))^2 \equiv l(l+1)C_l/2 \pi$. We do not discuss the
first peak position and height because they are almost unaffected by the
change in $Y_p$. We calculate the responses of these quantities with
respect to the change in the cosmological parameters around the fiducial
values, $\omega_m=0.14, \omega_b=0.024, \Omega_\Lambda=0.73, \tau
=0.166, n_s = 0.99$ and $Y_p=0.24$ where $\omega_i \equiv \Omega_i h^2$
with $\Omega_i$ being the energy density of component $i$ normalized by
the critical energy density. The subscript $b$ denotes baryon and $m$
stands for matter which is the sum of baryon and CDM. $h$ is the Hubble
parameter, $\tau$ is the reionization optical depth and $n_s$ is the
scalar spectral index of primordial power spectrum. These values (except
for $Y_p$) are the mean values of WMAP for the power-law $\Lambda$CDM
model \cite{Spergel:2003cb}. Also, we keep the flatness of the universe
when we change parameters. We found
\begin{eqnarray}
\Delta H_2 &=& 
-0.30\frac{\Delta \omega_b}{\omega_b} 
+0.015\frac{\Delta \omega_m}{\omega_m} 
+0.41\frac{\Delta n_s}{n_s} 
-0.0125 \frac{\Delta Y_p}{Y_p},
\label{eq:H2} \\
\Delta H_3 &=&
-0.18\frac{\Delta \omega_b}{\omega_b} 
+0.21\frac{\Delta \omega_m}{\omega_m} 
+0.56\frac{\Delta n_s}{n_s} 
-0.029 \frac{\Delta Y_p}{Y_p},
\label{eq:H3}
\end{eqnarray}
where we neglected the dependence on $\Omega_\Lambda$ and $\tau$ since
their coefficients are very tiny even if compared to that of $Y_p$. From
these expressions, we can see that the response of $C_l$ to the change
in $Y_p$ is very sluggish. This is one of the reasons why we do not
expect to obtain a meaningful constraint on $Y_p$ from CMB until
recently. Moreover, the change of $C_l$ caused by varying $Y_p$ is
readily canceled by shifting other parameters. However, since
observations of CMB now have become precise and cover wider multipole
range, we may have a chance to constrain $Y_p$ from current observations
of CMB, which will be discussed in the next section.

\section{Constraint on $Y_p$ from current CMB observations}

Now we study the constraint on the helium abundance from current
observations of CMB.  For this purpose, we use the data from WMAP
\cite{Bennett:2003bz}, CBI \cite{Readhead:2004gy} and ACBAR
\cite{Kuo:2002ua}.  We also include the recent data from BOOMERANG
experiment \cite{Jones:2005yb,Piacentini:2005yq,Montroy:2005yx}.  To
calculate $\chi^2$ from WMAP data, we used the code provided by WMAP
\cite{Kogut:2003et,Hinshaw:2003ex,Verde:2003ey}. For CBI, ACBAR and
BOOMERANG, we made use of modules in COSMOMC \cite{Lewis:2002ah}.  As
mentioned in the introduction, two groups have reported different bound
on $Y_p$ using CMB data alone, especially in terms of its uncertainties.
One group has obtained $ 0.160 \le Y_p \le 0.501 $ \cite{Trotta:2003xg}
at 1$\sigma$ C.L., on the other hand the authors of
Ref.~\cite{Huey:2003ef} give the bound as $Y_p=
0.250^{+0.010}_{-0.014}$. The authors of Refs.~\cite{Trotta:2003xg} and
\cite{Huey:2003ef} use the CMB data which cover the similar multipole
region as that of ours. For details of their analysis, we refer the
reader to Refs.~\cite{Trotta:2003xg} and \cite{Huey:2003ef}.  If the
severe bound on $Y_p$ is obtained from current CMB data, it means that
the CMB power spectrum is sensitive to the value of $Y_p$ and the prior
on $Y_p$ would affect the constraints on other cosmological
parameters. Thus it is important to check the bound independently.

\begin{figure}[t]
\begin{center}
\scalebox{0.9}{\includegraphics{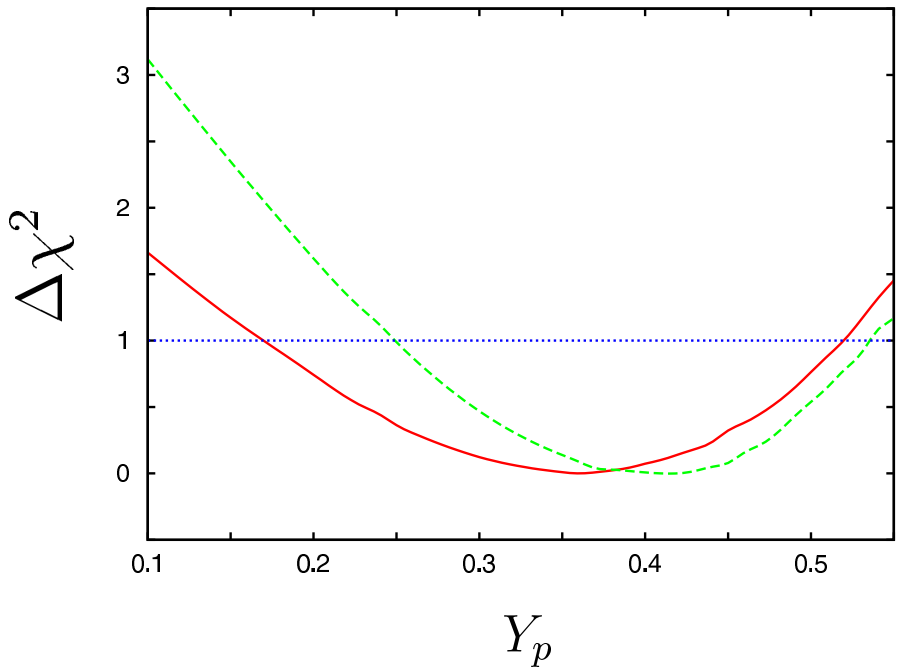}} \caption{The values of
$\Delta \chi^2$ are shown as a function of $Y_p$. Other cosmological
parameters are taken to minimize $\chi^2$. The red solid line is for the
data of WMAP, CBI and ACBAR. The green dashed line includes BOOMERANG
data in addition.}  \label{fig:current}
\end{center}
\end{figure}

For the analysis in this paper, we adopted a $\chi^2$ minimization by
nested grid search instead of Markov chain Monte Carlo (MCMC) method
which was used in their analysis.  In our analysis, we apply the Brent
method \cite{brent} of the successive parabolic interpolation to find a
minimum with respect to one specific parameter with other parameters at
a given grid, then we iteratively repeat the procedure to find the
global minimum. For the detailed description of this method, we refer
the readers to Ref.~\cite{Ichikawa:2004zi}. Here we assume a flat universe
and the cosmological constant for dark energy. We also assume no
contribution from gravity wave.  In Fig.~\ref{fig:current}, we show the
values of $\chi^2$ minimum as a function of $Y_p$. As seen from the
figure, we do not have a severe constraint from current observations of
CMB, which supports the result of Ref.~\cite{Trotta:2003xg}. Reading
$Y_p$ values which give $\Delta \chi^2 = 1$, we obtained the constraint
at 1$\sigma$ C.L. as $0.17 \le Y_p \le 0.52$ for the case where the data
from WMAP, CBI and ACBAR are used. When the data from BOOMERANG is
added, we got $0.25 \le Y_p \le 0.54$.
We have also made the analysis for different data sets for
comparison. In fact, we cannot obtain a significant constraint in the
region $0.1 < Y_p < 0.6$ using WMAP data alone. Even if we add the data
from BOOMERANG, we cannot constrain the value of $Y_p$.  Thus the data
from CBI and ACBAR which cover high multipole regions are important to
constrain $Y_p$ although the constraint is rather weak.

\begin{figure}[t]
\begin{center}
\scalebox{0.9}{\includegraphics{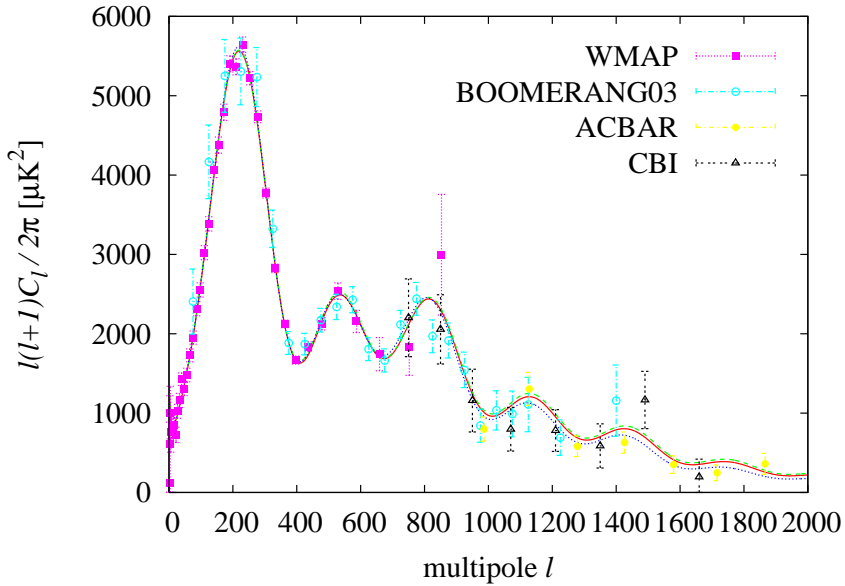}} \caption{The CMB power spectra
for the case with $Y_p=0.24, \omega_m=0.135,\omega_b=0.023,
h=0.72, \tau=0.117, n_s=0.96$ (red solid line), 
$Y_p=0.1, \omega_m=0.130, \omega_b=0.023,
h=0.72, \tau=0.101, n_s=0.95$ (green dashed  line) and 
$Y_p=0.5,\omega_m=0.148, \omega_b=0.023, h=0.72, \tau = 0.117, n_s=0.99$
(blue dotted  line).
Notice that these power
spectra are almost indistinguishable up to multipole region $l \sim
1000$.  } 
\label{fig:cmb2}
\end{center}
\end{figure}

As discussed in the previous section, the CMB power spectrum can be
affected by changing the value of $Y_p$. However this change can be
canceled by tuning other cosmological parameters to give almost the same
CMB power spectra.  To see this clearly, in Fig.~\ref{fig:cmb2}, we show
the CMB power spectra for several values of $Y_p$ with other
cosmological parameters being chosen to give almost indistinguishable
angular power spectra.  As seen from the figure, even if we take much larger
or smaller values of $Y_p$ than usually assumed, we can fit such values
of $Y_p$ to the data by tuning other cosmological parameters.

When we include the data from BOOMERANG, the favored values of $Y_p$ are
shifted to larger $Y_p$. Notice that higher multipoles are more
suppressed by increasing $Y_p$, which is almost the same effect as
decreasing $n_s$\footnote{
Eqs.~(\ref{eq:H2}) and (\ref{eq:H3}) show these properties
quantitatively. We remark that they are also useful to understand the 
tendency  that the values of $n_s$ which give the minimum $\chi^2$ for
fixed $Y_p$ become larger as we increase $Y_p$. This is because the
suppression of higher multipoles caused by increasing $Y_p$ can be
compensated by increasing $n_s$.
}.  Since BOOMERANG data favors red-tilted initial power spectrum
compared to other data such as WMAP \cite{MacTavish:2005yk}, it is
reasonable that larger values of $Y_p$ are favored by BOOMERANG.
Particularly, $Y_p=0.24$ which is used in usual analysis is just out of
the 1$\sigma$ bound. However, it is allowed at 2$\sigma$ C.L. so we do
not take this as a serious discrepancy from the standard assumption.

\section{Constraint on $Y_p$ from future CMB observations 
and the role of standard BBN theory}

In this section, we discuss the future constraint on the primordial
helium mass fraction and other cosmological parameters. We especially
want to investigate how the constraints are modified when we take
account of the relation between $\omega_b$ and $Y_p$ fixed by the
standard BBN theory.  As mentioned in the introduction, the CMB
anisotropies can be measured more precisely in the future, thus the
primordial helium mass fraction may well be determined from CMB
observations alone.  The future constraints on $Y_p$ has already been
investigated in Ref.~\cite{Trotta:2003xg} using the expected WMAP 4 year
data, Planck and cosmic variance limited experiments. Here we study this
issue supplementing the consideration regarding the prediction of the
BBN theory.

When we only consider observations of CMB alone, the primordial helium
mass fraction $Y_p$ can be viewed as one of free independent
parameters. However, when we take account of the BBN theory, $Y_p$ is
not an independent parameter any more but is related to the value of the
baryon-to-photon ratio or the energy density of baryon. Below, we
discuss how such relation derived from the BBN calculation affects the
determination of cosmological parameters in the future Planck
experiment.

First we give the relation between $Y_p$ and the baryon-to-photon ratio
$\eta$ from the calculation of the standard BBN. $\eta$ and the baryon
density $\omega_b$ are related as $10^{10} \eta = 273.49 \omega_b$. Some
groups have reported the fitting formula for $Y_p$ as a function of
$\eta$ \cite{Serpico:2004gx,Lopez:1998vk,Esposito:1999sz,Burles:2000zk}.
Here we adopt the fitting formula given in Ref.~\cite{Serpico:2004gx}
\begin{eqnarray}
10 Y_p &=& 
\left[
\sum_{n=1}^{8} a_n x^{n-1} 
+\sum_{n=1}^{8} b_n x^{n-1}  \Delta N
+\sum_{n=1}^{8} c_n x^{n-1}  \left( \Delta N \right)^2 
+\sum_{n=1}^{8} d_n x^{n-1}  \left( \Delta N \right)^3
\right] \notag \\ 
&& \times 
\exp \left( \sum_{n=1}^{6} e_n x^{n} \right),
\label{eq:Ypfit}
\end{eqnarray}
where $x \equiv \log_{10}( 10^{10} \eta)$, the coefficients $a_n, b_n,
c_n, d_n$ and $e_n$ are given in Ref.~\cite{Serpico:2004gx} and $\Delta N$
represents the number of effective degrees of freedom of extra
relativistic particle species. The standard BBN case is obtained with
$\Delta N = 0$. According to Ref.~\cite{Serpico:2004gx}, the accuracy of
this formula is better than 0.05 \% for the range of $ 5.48 \times
10^{-10} < \eta < 7.12 \times 10^{-10}$ ($0.02 < \omega_b < 0.026$)
which corresponds to the 3$\sigma$ range obtained from WMAP and
$-3<\Delta N<3$.  Since scenarios with $\Delta N \ne 0$ have been
discussed in the literature including the possibility of negative
$\Delta N$ such as dark radiation in brane world scenario, varying
gravitational constant and so on\footnote{
The negative values of $\Delta N$ can also arise in a scenario with low
reheating temperature $T_{\rm ref} \sim \mathcal{O}$(MeV). However, in
this kind of scenario, the neutrino distribution functions are deviated
from the thermal ones so the primordial helium abundance is modified in
a way that the fitting formula Eq.~(\ref{eq:Ypfit}) does not apply
\cite{Kawasaki:1999na,Kawasaki:2000en,Hannestad:2004px,Ichikawa:2005vw}. The
most recent $Y_p$ calculation in the low reheating scenario including
the effects of neutrino oscillations are given in
Ref.~\cite{Ichikawa:2005vw}.
}, we consider two cases when we discuss the future constraints.  The
first one is the case of the standard BBN, in other words, we assume
that the energy density of extra radiation component as a fixed
parameter with $\Delta N=0$.  For the other case, we treat $\Delta N$ as
a usual cosmological parameter which we vary, namely we assume $\Delta N
\ne 0$.  In this case, we use Eq.~(\ref{eq:Ypfit}) to obtain $Y_p$ for
given $\omega_b$ and $\Delta N$.

Now we discuss the expected constraint from future CMB observations.
For this purpose, we adopt the Fisher matrix method. Thus first we
briefly review the Fisher matrix analysis which is widely used in the
literature to study the future constraints on cosmological parameters.
Detailed descriptions of this analysis method can be found in
Refs.~\cite{Tegmark:1996bz,Jungman:1995av,Eisenstein:1998hr,Lesgourgues:2004ps}.
For the CMB data, the Fisher matrix can be written as
\begin{equation}
F_{ij} 
= \sum_l \sum_{X,Y} \frac{\partial C_l^X}{\partial x_i} 
{\rm Cov}^{-1} (C_l^X, C_l^Y) 
\frac{\partial C_l^Y}{\partial x_j},
\end{equation}
where $X,Y= TT, TE, EE$, $x_i$ represents a
cosmological parameter and $ {\rm Cov} (C_l^X, C_l^Y) $ is the
covariance matrix of the estimator of the corresponding CMB power
spectrum which is given explicitly in Ref.~\cite{Jungman:1995av}.  The
1$\sigma$ uncertainty can be estimated as $\sqrt{ (F^{-1})_{ii} }$ for a
cosmological parameter $x_i$.  For the fiducial model, we assumed the
cosmological parameters as $A=0.86$, $\omega_m=0.14, \omega_b=0.024,
\Omega_\Lambda=0.73, \tau =0.166$ and $n_s = 0.99$. Here, $A$ represents
the amplitude of scalar perturbation whose normalization is taken to be
same as that of WMAP team \cite{Spergel:2003cb}. The fiducial value for
$Y_p$ is fixed using the BBN relation Eq.~(\ref{eq:Ypfit}) for a given
$\omega_b$ and $\Delta N$ unless otherwise stated.  A flat universe is
assumed and we do not consider the contribution from the tensor
mode. For dark energy, we assumed the cosmological constant.  To
forecast uncertainties, we use the future data from Planck \cite{planck}
whose expected instrumental specifications can be found in
Ref.~\cite{Lesgourgues:2004ps}.

\begin{figure}[t]
\begin{center}
\scalebox{0.95}{\includegraphics{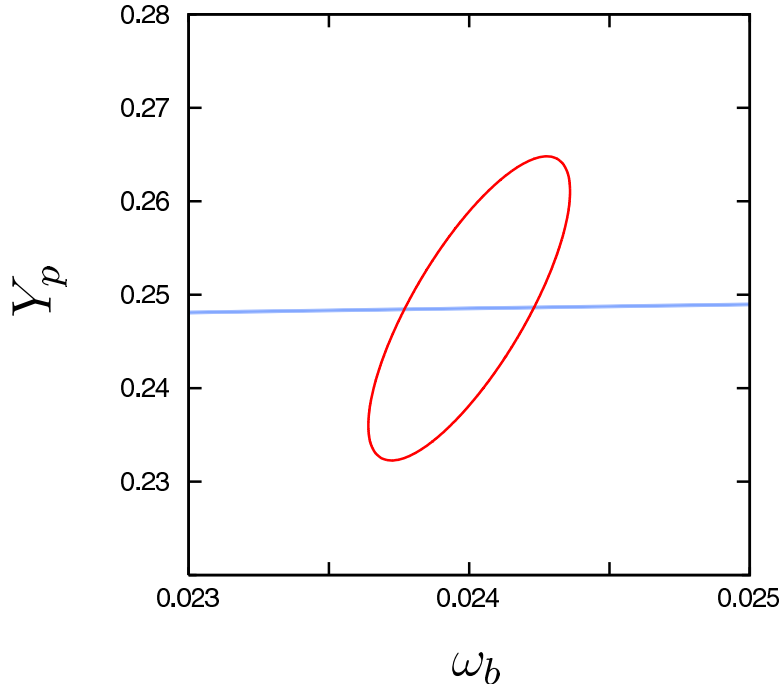}}
\caption{Expected contour of the 1$\sigma$ constraint from Planck
experiment. We take $Y_p$ as an independent free parameter. Other
cosmological parameters are marginalized over in this figure. The thin
and nearly horizontal band is the theoretical BBN calculation of $Y_p$
as a function of $\omega_b$ with its width representing 1$\sigma$ error
from reaction rates.
Here we assumed $\Delta N=0$.
}
\label{fig:future}
\end{center}
\end{figure}

Now we show our results. First we present the case with $Y_p$ being
treated as a free parameter, namely we do not consider the BBN relation.
In Fig.~\ref{fig:future}, we show the expected 1$\sigma$ contour in the
$\omega_b$ vs. $Y_p$ plane. Other cosmological parameters are
marginalized.  We also draw the band for $Y_p$ as a function of
$\omega_b$ from theoretical calculation of the standard BBN with
1$\sigma$ error due to the uncertainties in the reaction rates. The
uncertainty is dominated by that of the neutron lifetime, which is very
small. As is clearly seen from the figure, the constraint on $Y_p$ from
Planck is not significant compared to the uncertainties of the standard
BBN calculation. Hence as far as we take account of the standard BBN,
the helium mass fraction can be fixed using the BBN relation of
Eq.~(\ref{eq:Ypfit}) even for the precision measurements of CMB such as
Planck.

\begin{table}[t]
\begin{center}
\begin{tabular}{c|cccccccc}
\hline 
 & $A$ 
 & $\omega_b$ 
 & $\omega_m$ 
 & $\Omega_\Lambda$ 
 & $\tau$ 
 & $n_s$ 
 & $Y_p$ 
 & $\Delta N$ \\  \hline
 w/o BBN rel. $(\Delta N=0)$
 & $0.132    $ 
 & $0.00068 $  
 & $0.0033  $ 
 & $0.025   $  
 & $0.067   $ 
 & $0.025   $   
 & $0.030   $   
 & $-$          \\
 w/ BBN rel. $(\Delta N=0)$
 & $0.071    $ 
 & $0.00022  $  
 & $0.0018  $ 
 & $0.010   $  
 & $0.036  $ 
 & $0.005  $  
 &   $-$  
 &   $-$    \\ 
 w/o BBN rel. $(\Delta N \ne 0)$
 & $0.164    $ 
 & $0.00098 $  
 & $0.0041  $ 
 & $0.041   $  
 & $0.086   $ 
 & $0.038   $   
 & $0.031   $   
 & $0.41    $   \\ 
 w/ BBN rel.  $(\Delta N \ne 0)$
 & $0.120    $ 
 & $0.00074 $  
 & $0.0031  $ 
 & $0.035   $  
 & $0.065   $ 
 & $0.029   $ 
 & $-$ 
 & $0.41  $   \\ \hline
\end{tabular}
\caption{Expected 1$\sigma$ uncertainties from Planck experiment using
the temperature fluctuation alone. See the text for the fiducial values used in the
analysis. } \label{tab:Yp_nopol}
\end{center}
\end{table}

\begin{table}
\begin{center}
\begin{tabular}{c|ccccccccc}
\hline 
 & $A$ 
 & $\omega_b$ 
 & $\omega_m$ 
 & $\Omega_\Lambda$ 
 & $\tau$ 
 & $n_s$ 
 & $Y_p$ 
 & $\Delta N$ \\  \hline
 w/o BBN rel. $(\Delta N=0)$
 & $0.0106   $ 
 & $0.00024 $  
 & $0.0012  $ 
 & $0.0081  $  
 & $0.0054  $ 
 & $0.0075  $   
 & $0.011   $ 
 & $-$ \\
 w/ BBN rel. $(\Delta N=0)$
 & $0.0097  $ 
 & $0.00015 $  
 & $0.0012  $ 
 & $0.0067  $  
 & $0.0052  $ 
 & $0.0035  $ 
 & $-$
 & $-$   \\ 
 w/o  BBN rel. $(\Delta N \ne 0)$
 & $0.0107   $ 
 & $0.00025 $  
 & $0.0029  $ 
 & $0.0090  $  
 & $0.0055  $ 
 & $0.0082  $   
 & $0.014   $   
 & $0.19   $   \\ 
 w/ BBN rel. $(\Delta N \ne 0)$
 & $0.0099  $ 
 & $0.00023 $  
 & $0.0020  $ 
 & $0.0091  $  
 & $0.0054  $ 
 & $0.0078  $ 
 & $-$ 
 & $0.14  $ 
   \\ \hline
\end{tabular}
\caption{Expected 1$\sigma$ uncertainties from Planck experiment using
both the temperature and polarization data. See the text for the fiducial values used
in the analysis. } \label{tab:Yp_all}
\end{center}
\end{table}

Next we discuss how the theoretical BBN relation can affect the
determinations of cosmological parameters. First we consider the case
with $\Delta N=0$.  When we determine the value of $Y_p$ for a given
$\omega_b$ using Eq.~(\ref{eq:Ypfit}) (to be more specific, when we
calculate numerical derivatives with respect to $\omega_b$, we
simultaneously varied $Y_p$ following Eq.~(\ref{eq:Ypfit})) we can
expect that the uncertainties of other cosmological parameters are
reduced to some extent. In Tables \ref{tab:Yp_nopol} and
\ref{tab:Yp_all}, we show the uncertainties of cosmological parameters
from the future Planck experiment using the information of TT spectrum alone
and that including polarization spectrum, respectively.  The first and second rows in
the tables correspond to the case without and with the BBN relation. For
these cases, we assumed $\Delta N = 0$.  Furthermore, we also show the
cases with $\Delta N \ne 0$ in the third and forth row in the tables.

Now we discuss the cases with $\Delta N=0$. As seen from the tables,
when we assume the BBN relation, the uncertainties become smaller by a
factor of $\mathcal{O} (1)$ compared to that for the case with $Y_p$
being an independent free parameter. The parameter which receives the
benefit most is $n_s$. This is consistent with the fact that this
parameter is the most degenerate parameter with $Y_p$. Meanwhile, we
note that the value of $Y_p$ fixed by the BBN relation for
$\omega_b=0.024$ is slightly different from $Y_p=0.24$ which is usually
used in the literature. We also evaluated the uncertainties fixing the
helium mass fraction as $Y_p=0.24$ independent of $\omega_b$ and checked
that the 1$\sigma$ errors are quite similar to those for the case with
$Y_p$ being related to $\omega_b$ by Eq.~(\ref{eq:Ypfit}).  Since the
change of $C_l$ with respect to that of $Y_p$ is very small, in other
words the derivative of $C_l$ with respect to $Y_p$ is very small, we
can have almost the same result even if we use a slightly different
value for $Y_p$.  Thus we can just fix the value of $Y_p$ instead of
using Eq.~(\ref{eq:Ypfit}) even for the future CMB experiments such as
Planck.
 
\begin{figure}[t]
\begin{center}
\scalebox{0.95}{\includegraphics{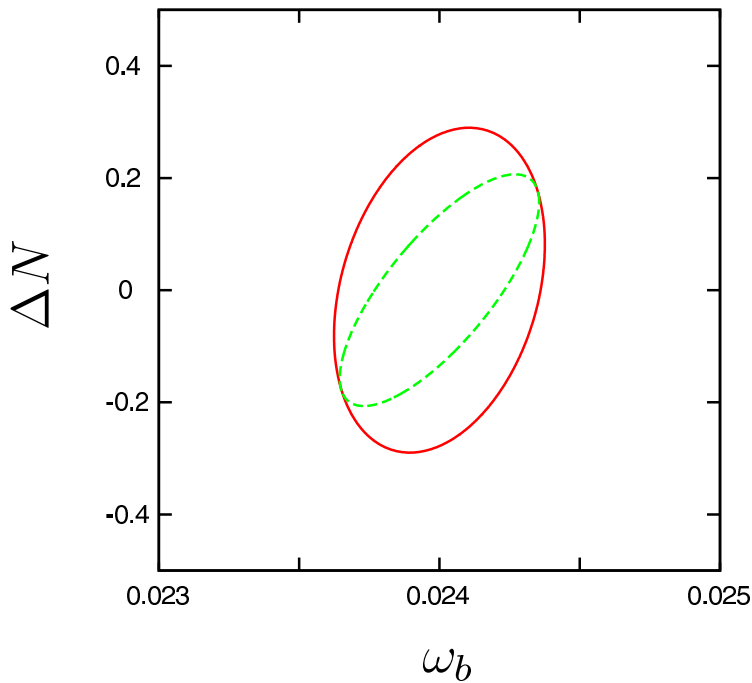}}
\caption{Expected contours of 1$\sigma$ constraints from Planck
experiment for the cases without the BBN relation (solid red line) and
with it (dashed green line). We marginalized over other cosmological
parameters in this figure. $Y_p$ is determined by the BBN relation.}
\label{fig:future2}
\end{center}
\end{figure}

Here the discussion for the case with $\Delta N \ne 0$ is in order. Here
we treat $\Delta N$ as one of the cosmological parameters which should
be varied.  Notice that the addition of an extra radiation component 
can affect the CMB power spectrum through the
speed up of the Hubble expansion and the early ISW effect because of the change of 
the radiation-matter equality epoch. 
 In fact, some authors have discussed the future constraints
on cosmological parameters paying attention to $\Delta N$
\cite{Lopez:1998aq,Bowen:2001in} but without considering the effect of
$Y_p$. Here we study this issue allowing $Y_p$ to vary and also
investigate the implications of the BBN relation on the future
constraints on them.  For the purpose of the Fisher matrix analysis, 
we assumed $\Delta N = 0$ as the fiducial value.
As already discussed, $Y_p, \omega_b$ and $\Delta
N$ are related by Eq.~(\ref{eq:Ypfit}) from the BBN calculation.  In
Fig.~\ref{fig:future2}, we show the expected 1$\sigma$ contour in the
$\omega_b$ vs. $\Delta N$ plane from Planck experiment for the cases
with and without assuming the BBN relation. Naturally, when we assume
the BBN relation, the uncertainties become smaller because it reduces
the number of independent parameters. Also notice that, since the BBN
theory relates $\omega_b$ with $\Delta N$, the contour shrinks to the
direction of correlation between $\omega_b$ and $\Delta N$.

In the third and forth rows of Tables \ref{tab:Yp_nopol} and
\ref{tab:Yp_all}, the uncertainties of cosmological parameters are also
shown for the cases with and without the BBN relation being imposed
among $Y_p,\omega_b$ and $\Delta N$ respectively. As expected, the
uncertainties for cosmological parameters for the case with the BBN
relation are smaller, however the differences are not so large.  

Here we comment on the implications of the BBN relation on the
constraints from current observations of CMB. We have also made the
analysis adopting the BBN relation to fix the value of $Y_p$ for given
$\omega_b$. The constraint on $\omega_b$ for this case is almost
unchanged compared to the case with $Y_p=0.24$ being fixed.  This can be
more or less expected from the result we have shown in the previous
section. This again shows that current observations of CMB are not so
sensitive to the values of $Y_p$. Thus predicting the helium abundance
by the BBN theory using the CMB value of $\omega_b$, namely the procedure adopted in Refs.~\cite{Cyburt:2003fe,Cuoco:2003cu,Coc:2003ce,Cyburt:2004cq,Serpico:2004gx}, is valid at least with 
the current quality of the CMB data.

\section{Summary}

We revisited the constraint on the primordial helium mass fraction $Y_p$
from current observations of CMB. Some authors have already studied the
constraint \cite{Trotta:2003xg,Huey:2003ef}, however their results were
different especially in terms of the uncertainty.  One of the main
purpose of the present paper is to study the constraints on $Y_p$ from
current observations adopting a different analysis method.  Instead of
MCMC method which was adopted by the authors of
Refs.~\cite{Trotta:2003xg,Huey:2003ef} to obtain the constraint, here we
adopted a $\chi^2$ minimization by a nested grid search. We did not
obtain a severe constraint in agreement with
Ref.~\cite{Trotta:2003xg}. Using the data from WMAP, CBI and ACBAR as
well as recent BOOMERANG data, we get 1$\sigma$ constraint as 
$ 0.25 \le Y_p \le 0.54$ and $0.17 \le Y_p \le 0.52$  for the cases with and
without the data from BOOMERANG. It might be interesting to note that
usual assumption of $Y_p=0.24$ is not in the 1$\sigma$ error range of
BOOMERANG combined analysis but it is not of high significance at this
stage so we can safely assume $Y_p=0.24$ for current CMB data analysis.

We also studied the future constraint from CMB on $Y_p$ taking account
of the standard BBN prediction as a prior on $Y_p$.  Although we cannot
obtain a severe constraint at present, observations of CMB can be much
more precise in the future. Thus we may have a chance to obtain a
precise measurement of $Y_p$ from upcoming CMB experiments. On the other
hand, since the primordial helium has been formed during the time of
BBN, once the baryon-to-photon ratio is given, the value of $Y_p$ can be
evaluated theoretically assuming the standard BBN. Thus, in this case,
we do not have to assume $Y_p$ as an independent free parameter when we
analyze CMB data.  We studied how such BBN theory prior on $Y_p$ affects
the determination of other cosmological parameters in the future Planck
experiment. We evaluated the uncertainties for the case with $Y_p$ being
an independent free parameter and $Y_p$ being fixed for a given
$\omega_b$ using the BBN relation.  We showed that the BBN prior
improves the constraints on other cosmological parameters by a factor of
$\mathcal{O}(1)$ and also it induces some correlations among the
parameters which appear in the BBN relation.  As shown in
Fig.~\ref{fig:future}, as far as we consider the standard scenario of
cosmology, the helium mass fraction can be fixed for CMB analysis even
in the future experiments since we can expect that the constraint from
Planck is much weaker than the uncertainty of the theoretical
calculation of the standard BBN.  However, it is worthwhile to do CMB
analysis treating $Y_p$ as a free parameter and measure the helium mass
fraction independently from the baryon density since it provides a
consistency test for the standard BBN theory (of course, measurements of
primordial light element abundances by astrophysical means provide
further consistency tests).  By checking the robustness of the
consistency from various observations, the golden age of precision
cosmology can push us toward the accurate understanding of the universe.

\bigskip 
\noindent 
{\bf Acknowledgment:} We acknowledge the use of
CMBFAST \cite{Seljak:1996is} package for our numerical calculations.
The work of T.T. is supported by Grand-in-Aid for JSPS fellows.

\end{document}